\documentclass[twocolumn,english,showpacs,footinbib,aps,prb]{revtex4}

\usepackage[normalem]{ulem}

\usepackage[T1]{fontenc}
\usepackage[latin9]{inputenc}
\usepackage{graphicx}
\usepackage{amsmath}
\usepackage{amssymb}
\usepackage{bbm}
\usepackage{xspace}
\usepackage{color}
\usepackage{footnote}
\usepackage{comment}
\usepackage{hyperref}
\usepackage{subfigure}

\usepackage{ulem}
\definecolor{grey}{RGB}{100,100,100}

\definecolor{green2}{RGB}{00,100,00}

\definecolor{dgreen}{rgb}{0.0,0.5,0.0}

\definecolor{orange}{RGB}{252,77,6}
\definecolor{brown}{RGB}{200,127,50}
\definecolor{blue}{RGB}{00,000,100}
\definecolor{blue2}{RGB}{00,000,250}
\definecolor{green1}{RGB}{00,100,00}
\definecolor{green2}{RGB}{00,150,00}
\definecolor{green3}{RGB}{00,200,00}
\definecolor{green4}{RGB}{00,250,00}

\newcommand{\locGFRinv}[1]{[g^{-1}_{#1}]^{R}}
\newcommand{\locGFKinv}[1]{[g^{-1}_{#1}]^{K}}
\newcommand{\locGF}[1]{g^{\gamma}_{#1}}
\newcommand{\kxy}{{{\bf k} }}

\newcommand{\beq}{\begin{equation}}
\newcommand{\eeq}{\end{equation}}

\newcommand{\Imm}{{\rm Im\:}}

\begin{document}

\title{Resonance Effects in Correlated Multilayer Heterostructures}

\author{Irakli Titvinidze}

\email{irakli.titvinidze@tugraz.at}


\affiliation{Institute of Theoretical and Computational Physics, Graz University
of Technology, 8010 Graz, Austria}

\author{Antonius Dorda}

\affiliation{Institute of Theoretical and Computational Physics, Graz University
of Technology, 8010 Graz, Austria}

\author{Wolfgang von der Linden}

\affiliation{Institute of Theoretical and Computational Physics, Graz University
of Technology, 8010 Graz, Austria}

\author{Enrico Arrigoni}

\affiliation{Institute of Theoretical and Computational Physics, Graz University
of Technology, 8010 Graz, Austria}


\begin{abstract} 
We study the occurrence of negative differential conductance induced by  resonance effects in a model for a  multilayer heterostructure. In particular, we consider a system consisting of several correlated and 
non-correlated monoatomic layers, sandwiched between two metallic leads. The geometry confines electrons in wells within the heterostructures, which are connected to each other and to the leads by tunneling 
processes. The non-equilibrium situation is produced by applying a bias-voltage to the leads. Our results show that for specific values of the parameters  resonance tunneling  takes place. We  investigate in 
detail its influence on the current-voltage characteristics. Our results are obtained via non-equilibrium real-space dynamical mean-field theory. As an impurity solver we use the so-called auxiliary master 
equation approach, which addresses the impurity problem within an auxiliary system consisting of a correlated impurity, a small number of uncorrelated bath sites, and two Markovian environments described by 
a generalized master equation. 
\end{abstract}

\ifx\clength\undefined
\maketitle
\else
\nocomm
\fi

\begin{figure}[t]
\includegraphics[width=\columnwidth]{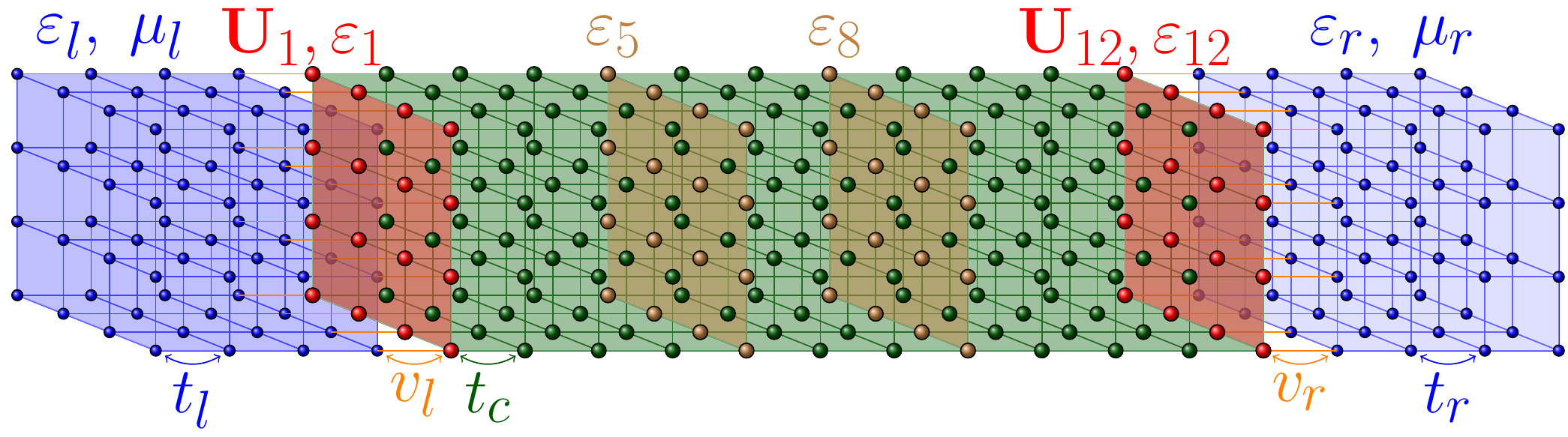}
\caption{(Color online) 
Schematic representation of the triple-well system studied in this paper, consisting of a central region of  12 layers sandwiched between two semi-infinite metallic leads (blue), with chemical potentials 
$\mu_{l/r}$ and onsite energies $\varepsilon_{l/r}=\mu_{l/r}$, respectively. A Hubbard interaction is only present in the boundary layers (red, $z=1,12$) with interaction strength $U_{1}=U_{12}=U=8$ and onsite 
energies $\varepsilon^{(0)}_{1}=\varepsilon^{(0)}_{12}=-U/2$. They form the outer walls of the quantum wells. The inner walls are  the  layers 5 and 8 (brown) and are caused by discontinuous onsite energies 
$\varepsilon_5^{(0)}=-\varepsilon_8^{(0)}=-4$, while all other layers have  $\varepsilon^{(0)}_{z}=0$.  Energies are measured in units of the  nearest neighbor hopping $t_c$ within the central region. For  
the nearest neighbor hopping within the leads we use $t_l=t_r=2$, and the hybridization between the left (right) lead and the  central region is $v_l=v_r=1$. A bias voltage $\Phi:=\mu_{l}-\mu_{r}$ is applied 
to the leads, which  linearly shifts the onsite energies of each layer: ${\varepsilon_z =\varepsilon_z^{(0)}+\mu_{l}- z\;(\mu_{l}-\mu_{r})/(L+1)}$}.
\label{schematicp}
\end{figure}

\section{Introduction}

Quantum mechanical resonance effects   play an important role in physics and technology. A well known   example is resonant tunneling through potential barriers. Tunneling through two barriers, which becomes 
resonant at a specific external bias voltage, underlies the functioning of resonant tunneling diodes.  Their applications range from high-speed microwave systems to novel digital logic circuits. 
Resonant tunneling through potential barriers is interesting from the theoretical point of view as well. To investigate this effect, one usually  considers double  
or multi-well structures made of semiconductor\cite{ch.es.74, datta.ch2-4-5, gu.li.96, er.po.11} or hybrid superconductor-semiconductor\cite{gi.pi.01} materials, 
graphene\cite{ro.dr.16, yo.ki.09, la.wa.15, fe.je.12} and graphene-boron\cite{gu.to.16, fa.le.15, br.lo.15, ba.fe.15, mi.tu.14, brey.14, ba.ga.14} heterostructures. Different approaches are used 
to theoretically investigate their properties. One can mention, for example,  modified optical Bloch equations\cite{gu.li.96}, self-consistent non-equilibrium Green's functions,\cite{er.po.11, go.sh.16u}
the  envelope wave-function formalism,\cite{go.sh.16u}  adiabatic approximations\cite{pr.sj.96}, combinations of   quantum transport random matrix theory  with  Bogoliubov-de Gennes equations\cite{gi.pi.01}, 
first-principle density functional theory\cite{br.lo.15}, Bardeen transfer Hamiltonian approach\cite{fe.je.12}, Wentzel-Kramers-Brillouin\cite{ro.dr.16}, and Lorentzian approximation for the quasi-particle 
spectral function\cite{gu.to.16}. However, to our knowledge, effects of electron correlations on resonant tunneling have so far been either neglected or included in a perturbative or mean-field way only.
Here, we  present a first study which examines the effect of correlations on resonant tunneling in an accurate and non-perturbative manner.

Recent experimental progress makes it possible to fabricate correlated 
heterostructures\cite{an.ga.99, is.og.01,ga.ah.02, oh.mu.02, oh.hw.04, zh.wa.12} with atomic resolution and in particular, growing atomically  abrupt layers with different electronic structures\cite{is.og.01, 
oh.mu.02, ga.ah.02}. Here, we study a system which is composed of alternating strongly correlated and non-correlated metallic layers, as well as band insulator layers (see Fig.~\ref{schematicp}). 
The geometry of the system is such that electrons are confined in three  wells connected by tunneling. The non-equilibrium situation is driven by applying a bias-voltage to the leads, which introduces 
a homogeneous electric field in the central region.  
Resonant tunneling is mainly induced by the particular geometry, rather than the specific values of the system parameters. Since our goal is to investigate the qualitative behavior of this effect,  
we mainly perform calculations for one representative set of model parameters. In addition, in order to address the effect of correlations on resonance tunneling, we also investigate the behavior of the 
resonance current as a function of the interaction $U$.

In contrast to the previous works mentioned above, we use dynamical mean-field theory (DMFT)\cite{ge.ko.96, voll.10, me.vo.89}, which can treat electron-electron correlations accurately and is one 
of the most powerful methods to investigate high-dimensional correlated systems. Originally, DMFT  was developed to treat equilibrium situations, and later  extended\cite{ao.ts.14, sc.mo.02u, fr.tu.06, free.08, 
jo.fr.08, ec.ko.09, okam.07, okam.08, ar.kn.13, ti.do.15, do.ti.16} to the  nonequilibrium case. This is formulated within the nonequilibrium Green's function approach originating from the works of 
Kubo\cite{kubo.57}, Schwinger\cite{schw.61}, Kadanoff, Baym\cite{ba.ka.61, kad.baym} and Keldysh\cite{keld.65}.

DMFT is a comprehensive, thermodynamically consistent and non-perturbative scheme which becomes exact in infinite dimensions, but usually quite well describes two and three dimensional systems. The only 
approximation in DMFT is locality of the self-energy. The latter can be calculated by mapping the original problem onto a single impurity Anderson model (SIAM)\cite{ande.61}, whose  parameters are determined 
self-consistently. For  homogeneous systems  the self-energies are the same for each lattice site due to  translational symmetry,  and, therefore, one needs to solve only one SIAM problem. For systems with broken 
translational invariance,  as in the present case, the self-energies depend on the layer index $z$. Therefore, it is necessary to generalize the  formalism and take into account the spatial 
inhomogeneity of the system\cite{po.no.99.sm, po.no.99.ms, po.no.99.ldmft, po.no.99.emfl, free.06, no.ma.10, is.li.09, no.ma.11, okam11, mi.fr.01, free.04, ok.mi.04.ldmft, do.ko.97, do.ko.98, so.yu.08, we.ha.11u, he.co.08, 
ko.hi.08, ko.hi.09,  no.ka.09,  ko.ba.11, bl.go.11, ki.ki.11, au.as.15, sn.ti.08, sn.ti.11, ti.sc.12, go.ti.10, sc.ti.13, au.ti.15, ze.fr.09, okam.07, okam.08, ec.we.13, ec.we.14, ha.fr.11}, and, 
accordingly to solve several SIAM problems.

In the present work  the nonequilibrium SIAM problem  is treated by using a recently developed auxiliary master equation approach\cite{ar.kn.13,do.nu.14,ti.do.15}, which treats the impurity problem within an 
auxiliary system consisting of a correlated impurity, a small number of uncorrelated bath sites and two Markovian environments described by a generalized master equation.

The paper is organized as follows: Sec.~\ref{Model} we introduce the Hamiltonian of the system. In Sec.~\ref{Method} we illustrate the application of  real-space dynamical mean-field theory within the 
non-equilibrium steady-state  Green's function  formalism for a system consisting of many layers. Afterwards, in Sec.~\ref{Results}, we present our results. Our conclusions are presented in  
Sec. \ref{Conclusions}.

\section{Model}\label{Model}

The model, consisting of a central region ($c$) with $L=12$ infinite and translationally invariant layers sandwiched between two semi-infinite metallic leads ($\alpha=l,r$), is described by the Hamiltonian 
(see Fig.~\ref{schematicp}):
\begin{eqnarray}
\label{Hamiltonian}
{\cal H}&=&
\hspace{-0.15cm}
-\hspace{-0.5cm}\sum_{z, \langle {\bf r}^{\phantom\dagger}_\perp,{\bf r}'_\perp\rangle_z, \sigma}\hspace{-0.5cm}t_{z} c_{z,{\bf r}_\perp,\sigma}^\dagger c_{z,{\bf r}'_\perp,\sigma}^{\phantom\dagger} 
-\hspace{-0.45cm}\sum_{\langle z, z'\rangle, {\bf r}_\perp, \sigma}\hspace{-0.45cm} t_{zz'} c_{z,{\bf r}_\perp,\sigma}^\dagger c_{z',{\bf r}_\perp,\sigma}^{\phantom\dagger} \nonumber \\
&+&\hspace{-0.1cm}\sum_{z,{\bf r}_\perp}\hspace{-0.1cm} U_z n_{z,{\bf r}_\perp,\uparrow}n_{z,{\bf r}_\perp,\downarrow} 
+\hspace{-0.1cm}\sum_{z,{\bf r}_\perp,\sigma}\hspace{-0.1cm}\varepsilon_z n_{z,{\bf r}_\perp,\sigma} \, ,
\end{eqnarray}
with nearest-neighbor  inter-layer (intra-layer) hopping $t_{zz'}$ ($t_{z}$), local onsite Hubbard interaction $U_z$ and local energy $\varepsilon_z$. $\langle z, z'\rangle$ stands for neighboring $z$ and $z'$ 
layers and $\langle {\bf r}^{\phantom\dagger}_\perp,{\bf r}'_\perp\rangle_z$ stands for neighboring ${\bf r}_\perp$ and ${\bf r}'_\perp$ sites of the $z$-th layer. $c_{z,{\bf r}_\perp,\sigma}^\dagger$ 
creates an electron at site ${\bf r}_\perp$  of  layer $z$ with spin ${\sigma}$ and 
$n_{z, {\bf r}_\perp, \sigma}=c_{z, {\bf r}_\perp, \sigma}^\dagger c_{z, {\bf r}_\perp, \sigma}^{\phantom\dagger}$ denotes the corresponding occupation-number operator. $z=1, \ldots, 12$ describes 
the central layers, while $z<1$ and $z>12$ corresponds to the left and the right lead layers, respectively. 

We assume isotropic nearest-neighbor hopping parameters  within the central region ($t_{zz'}=t_{z}=t_c$) and within  the  leads ($t_{zz'}=t_{z}=t_{\alpha=l,r}$). The hybridization between the leads and central 
region is the same on both sides $t_{0,1}=v_{l}=t_{12,13}=v_{r}$.

Finally, the local energy and the chemical potential in the leads is determined by an applied voltage $\Phi$, i.e. \hbox{$\varepsilon_{z<1}=\mu_{l} =  \Phi/2$} and \hbox{$\varepsilon_{z>12}= \mu_{r} = -\Phi/2$}.

The leads are initially prepared in equilibrium and $T=0$ at the
distant past (time $ \to -\infty$) when the hoppings between leads and layer are switched off. Then the hoppings are switched on and the system 
is allowed to evolve in time until steady state is reached. 
Notice that despite of the appearance of equilibrium Green's functions
\eqref{galphaK} in the expressions, 
 there is no approximation of fixing the leads in equilibrium.
 In our approach, 
it is not necessary to solve explicitly for the transient time evolution, and we can directly address the steady state. Since the leads are infinite, they 
have equilibrium properties far away from the device, but near the
device (within the healing length) there will be charge depletion or
enhancement, i.e. charge reconstruction near the interfaces.
In combination with the long-range part of the Coulomb interaction (LRCI)
this could induce modifications in  the singls-particle potential.
LRCI could be included by a simultaneous solution of the Poisson and DMFT
equation (see, e.g. \cite{free.06}), but this is beyond the scope of the present paper.
Notice that 
this approximation is common in the framework of real-space DMFT calculations  (see e.g. Refs.~\cite{okam.07,okam.08,kn.li.11, 
ne.ar.15, ma.am.15, ma.am.16, am.we.12, ri.an.16, ze.fr.09, ec.we.13, ec.we.14, ha.fr.11, po.no.99.sm, po.no.99.ms, po.no.99.ldmft, po.no.99.emfl}).
Here,  we approximate the effects of the LRCI, by introducing a linear behavior of the onsite energies (homogeneous electric field) in the central region as 
\hbox{${\varepsilon_z =  \varepsilon_z^{(0)} + \mu_{l} - z \Phi /(L+1)}$}.

\section{Real-space Dynamical Mean-Field theory }\label{Method}

In order to investigate steady-state properties we use real-space Dynamical mean-field theory (R-DMFT), which is also known as inhomogeneous DMFT. Due to the finite number of layers translational 
invariance along the $z$ axes (perpendicular to the layers) is broken, but the system  is still translationally invariant in the $xy$ plane. 
Therefore we can  introduce a corresponding momentum $\kxy=(k_x,k_y)$.~\cite{okam}

The Green's function for the  central region, can be expressed via Dyson's equation
\begin{eqnarray}
\label{GR}
&&[{\bf G^{-1}}]^{\gamma}(\omega, \kxy) =[{\bf g}_0^{-1}(\omega, \kxy)]^\gamma - {\boldsymbol \Sigma}^\gamma(\omega)-{\boldsymbol \Delta}^\gamma(\omega, \kxy) \hspace{0.5cm} \;.
\end{eqnarray} 

Here, we use boldface symbols to indicate matrices in the indices $z=1, \ldots, 12$. Moreover, $\gamma\in\{R,A,K\}$ stands for retarded, advanced and Keldysh components, respectively, 
and ${\bf G}^{A}(\omega, \kxy)=[{\bf G}^{R}(\omega, \kxy)]^{\dagger}$. 

The inverse of the non-interacting Green's function  for the isolated central region reads 
\begin{align} 
[{\bf g^{-1}_0}]^R_{zz'} (\omega, \kxy) &=\left(\omega + i 0^{+}-E_z( \kxy) \right)\delta_{zz'} + t_{zz'} \, ,
\label{g0R}\\
[{\bf g^{-1}_0}]^K_{zz'} (\omega, \kxy)&\simeq 0\;.
\end{align} 
with $E_z( \kxy)=\varepsilon_z -2t_z (\cos k_x  + \cos k_y)$.  ${\boldsymbol \Delta}^\gamma(\omega, \kxy)$ describes the hybridization with the leads and can be expressed as
\begin{equation}
\label{Delta}
{\boldsymbol \Delta}^\gamma_{zz'}(\omega, \kxy)=\delta_{z,z'}\left(\delta_{z,1}v_l^2 g_l^\gamma (\omega, \kxy)+ \delta_{z,L}v_r^2 g_r^\gamma (\omega, \kxy)\right) \, ,
\end{equation}
where $g_l^\gamma (\omega, \kxy)$ and $g_r^\gamma (\omega, \kxy)$ describe 
the  Green's functions for the edge layers of the leads disconnected from the central region. Their retarded 
component can be  expressed as\cite{po.no.99.sm, po.no.99.ms, hayd.80}
\begin{eqnarray}
&&g_\alpha^R(\omega, \kxy)= \frac{\omega -E_\alpha(\kxy)}{2t_\alpha^2}  
- i\frac{\sqrt{4t_\alpha^2-(\omega-E_\alpha(\kxy))^2}}{2t_\alpha^2} \, ,
\label{galphaR}
\end{eqnarray}
with $E_\alpha( \kxy)=\varepsilon_\alpha-2t_\alpha (\cos k_x  + \cos k_y)$. The sign of the square-root for negative argument must be chosen such that the Green's function has the correct $1/\omega$ 
behavior for $|\omega|\to \infty$. Since the disconnected leads are separately in equilibrium, we can obtain their Keldysh components from the retarded ones via the fluctuation dissipation theorem\cite{ha.ja}
\begin{equation}
g_\alpha^K(\omega, \kxy)= 2i(1-2f_\alpha(\omega))\;\Imm g_{\alpha}^R(\omega,\kxy) \, .
\label{galphaK}
\end{equation}
Here, $f_\alpha(\omega)$ is the Fermi distribution for chemical potential $\mu_\alpha$ and temperature $T_\alpha$.

Finally the self-energy ${\boldsymbol\Sigma}^\gamma_{zz'}(\omega)=\delta_{zz'} \Sigma^\gamma_z(\omega)$ is a diagonal and $\kxy$-independent matrix due to the DMFT approximation.
To determine the self-energy for each correlated layer $z$ we solve a (non-equilibrium) quantum impurity model with Hubbard interaction $U_z$ and onsite energy $\varepsilon_z$, coupled to a 
self-consistently determined bath. The latter is specified by its hybridization function obtained as (see e.g. Ref.~\cite{voll.10})
\begin{eqnarray}
&&\hspace{-0.5cm}\Delta_{{\rm bath},z}^R(\omega)=\omega + i 0^{+}-\varepsilon_z -\Sigma^R_z(\omega) -\frac{1}{G_{{\rm loc},z}^R(\omega)} \, , \\
\label{DeltaR} 
&&\hspace{-0.5cm}\Delta_{{\rm bath},z}^K(\omega)=- \Sigma^K_z(\omega)+\frac{G_{{\rm loc},z}^K(\omega)}{|G_{{\rm loc},z}^R(\omega)|^2} \
\label{DeltaK} 
\end{eqnarray}
where the local Green's function is defined as
\begin{equation}
G_{{\rm loc},z}^\gamma(\omega)=\int\limits_{\rm BZ} \frac{d^2{\kxy}}{(2\pi)^2} {\bf G}_{zz}^\gamma(\omega,\kxy) \;.
\label{G_loc}
\end{equation}
To calculate the  diagonal elements of the matrices ${\bf G}^\gamma(\omega,\kxy)$ one could invert the matrices in Eqs. \eqref{GR}. However, it is numerically more efficient to use the recursive Green's function 
method\cite{th.ki.81, le.ca.13, ec.we.13}, which we here  generalize to  Keldysh Green's functions. 
For a given $z$ we decompose the system into three decoupled clusters by setting $t_{z-1,z}=t_{z,z+1}=0$ (for the first and the last layer into two decoupled cluster). The result is  an isolated layer of the central 
region at position  $z$  and the two remaining parts of the central region to the left and to the right of layer $z$. By $L_{z-1}^\gamma(\omega,\kxy )$ ($R_{z+1}^\gamma(\omega,\kxy )$) we denote the  local Green's 
function at layer $z-1$ ($z+1$) of the isolated cluster to the left (right) of layer $z$. In addition, we define $\locGF{z}(\omega,\kxy )$ as the full cluster Green's function of layer $z$.\cite{inv_Keldish} 
For $z=2,\ldots,L-1$ it describes isolated layers, while for $z=1$ ($z=L$) it also contains the hybridization effects of the left (right) lead, which are covered by $\Delta^\gamma(\omega ,\kxy)$. For the sake 
of better readability, we will suppress the argument $(\omega ,\kxy)$ in the following equations. From \eqref{GR} and the  ensuing definitions we readily see that the inverse cluster Green's function 
$[g^{-1}_{z}]^{\gamma}$ is equal to diagonal elements of the inverse  $[{\bf G^{-1}}]^\gamma_{zz}$ of the full Green's function of the central region. The omitted hopping processes $t_{z-1,z}$ and $t_{z,z+1}$ 
can now be reintroduced by the Dyson equation, which is applicable due to the DMFT approximation of local self energies. We obtain
\begin{eqnarray}
\label{GRzz}
&&{\bf G}_{zz}^R=\frac{1}{\locGFRinv{z} - t_{z-1,z}^2 L_{z-1}^R -t_{z,z+1}^2 R_{z+1}^R} \, ,\\
\label{GKzz}
&&{\bf G}_{zz}^K=-\frac{\locGFKinv{z}  - t_{z-1,z}^2 L_{z-1}^K -t_{z,z+1}^2 R_{z+1}^K}{\bigl|\locGFRinv{z} - t_{z-1,z}^2 L_{z-1}^R -t_{z,z+1}^2 R_{z+1}^R \bigl|^2} \, .
\end{eqnarray}
The Green's functions $L_{z}^\gamma$ and $R_{z}^\gamma$ in turn are evaluated recursively as follows:
\begin{eqnarray}
\label{LR}
&&L_{z}^R=\frac{1}{\locGFRinv{z}- t_{z-1,z}^2 L_{z-1}^R} \, ,\\
\label{LK}
&&L_{z}^K=-\frac{\locGFKinv{z}- t_{z-1,z}^2 L_{z-1}^K}{\bigl|\locGFRinv{z} - t_{z-1,z}^2 L_{z-1}^R \bigl|^2}\, ,
\end{eqnarray}
for $z=2,3,\ldots L$ with initial values
 \begin{align}
\label{L_{1}}
L_{1}^R&=\frac{1}{\locGFRinv{1}} \, ,\qquad 
L_{1}^K=-\frac{\locGFKinv{1}}{\bigl|\locGFRinv{1}\bigl|^2} \, , 
\end{align}
and 
\begin{eqnarray}
\label{RR}
&&R_{z}^R=\frac{1}{\locGFRinv{z}- t_{z,z+1}^2 R_{z+1}^R} \, ,\\
\label{RK}
&&R_{z}^K=-\frac{\locGFKinv{z}- t_{z,z+1}^2 R_{z+1}^K}{\bigl|\locGFRinv{z} - t_{z,z+1}^2 R_{z+1}^R \bigl|^2} \, ,
\end{eqnarray}
for $z=L-1,L-2,\ldots 1$  
with initial values
\begin{align}
\label{R_L}
R_{L}^R&=\frac{1}{\locGFRinv{L}} \, ,\qquad
R_{L}^K=-\frac{\locGFKinv{L}}{\bigl|\locGFRinv{L}\bigl|^2} \, .
\end{align}

In addition, the self-consistent DMFT loop works as follows: we start with an initial guess for the self-energies $\Sigma^\gamma_z(\omega)$, which typically was taken equal to zero, and based 
on Eqs. \eqref{GR}-\eqref{G_loc} we calculate the bath hybridization functions $\Delta_{{\rm bath},z}^R(\omega)$ and $\Delta_{{\rm bath},z}^K(\omega)$ for each correlated site. From  them we solve the 
(non-equilibrium) quantum impurity models and calculate new self-energies as described below. We repeat this procedure until convergence is reached.\cite{convergance}

To address the impurity problem and evaluate self-energies, we adopt a recently developed auxiliary master equation approach (AMEA)\cite{ar.kn.13, do.nu.14, ti.do.15}. This method can be seen as  a generalization 
of the equilibrium exact-diagonalization impurity solver to treat  nonequilibrium steady-state situations. In AMEA dissipation, which is crucial in order to achieve a steady state, is included by additionally 
coupling the cluster to  Markovian environments, which can be seen as  particle sinks and reservoirs (for details see Refs. \onlinecite{ar.kn.13, do.nu.14, ti.do.15, do.ga.15}). The accuracy of the impurity 
solver increases with increase of $N_b$ and becomes exponentially exact in the limit $N_b \rightarrow \infty$.

\begin{figure}[t]
\includegraphics[width=0.75\columnwidth]{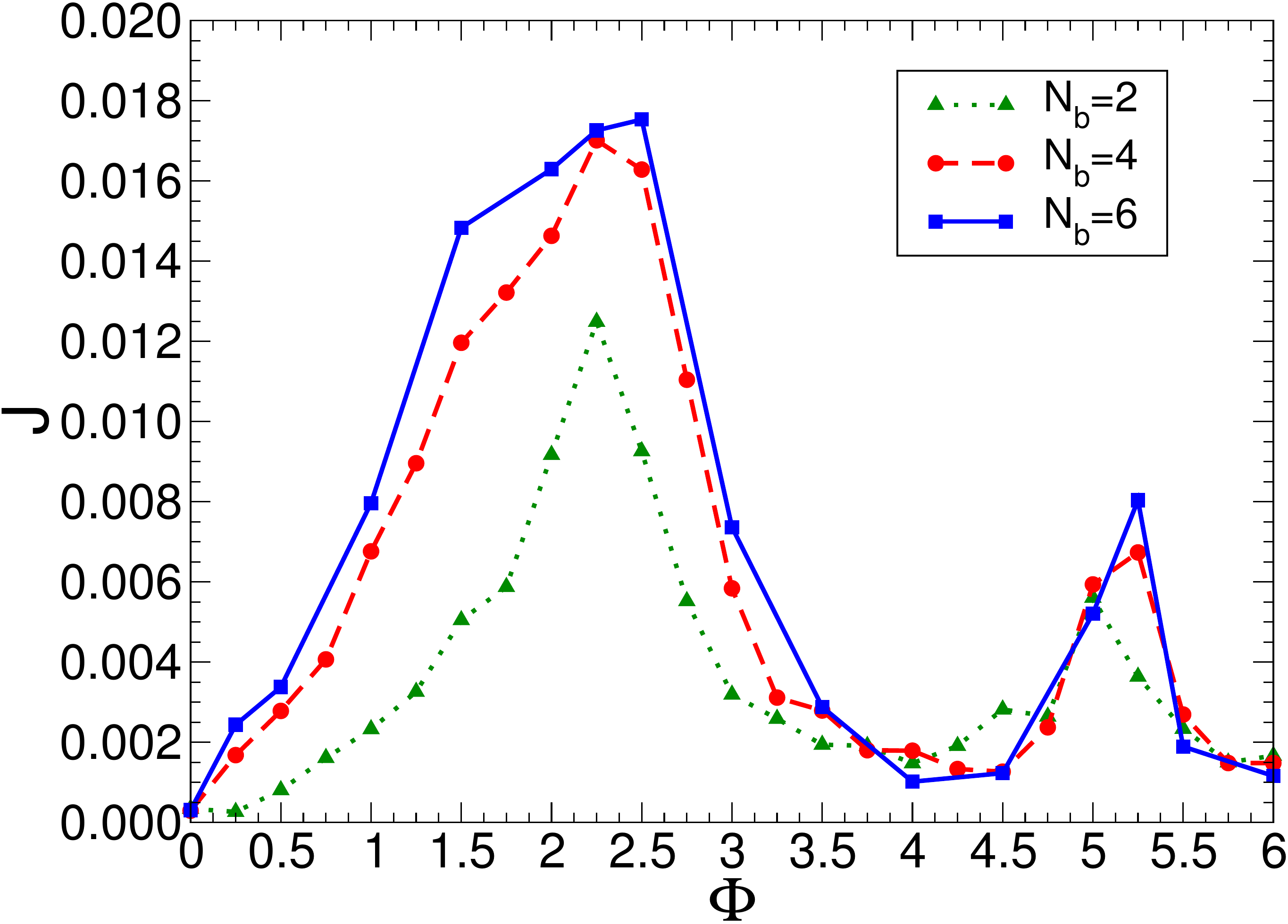}
\caption{(Color online) Current $J$ vs bias voltage $\Phi$. Solid, dashed and dotted lines  are obtained by solving the impurity problem with 
 $N_b=6$, $N_b=4$ and $N_b=2$, respectively (see text). 
Parameters are the same as in Fig.~\ref{schematicp} .
}
\label{Current_vs_Phi}
\end{figure}

\begin{center}
\begin{figure*}[t!]
\subfigure[]{
\label{Phi1}
\begin{minipage}[b]{0.45\textwidth}
\centering \includegraphics[width=1\textwidth]{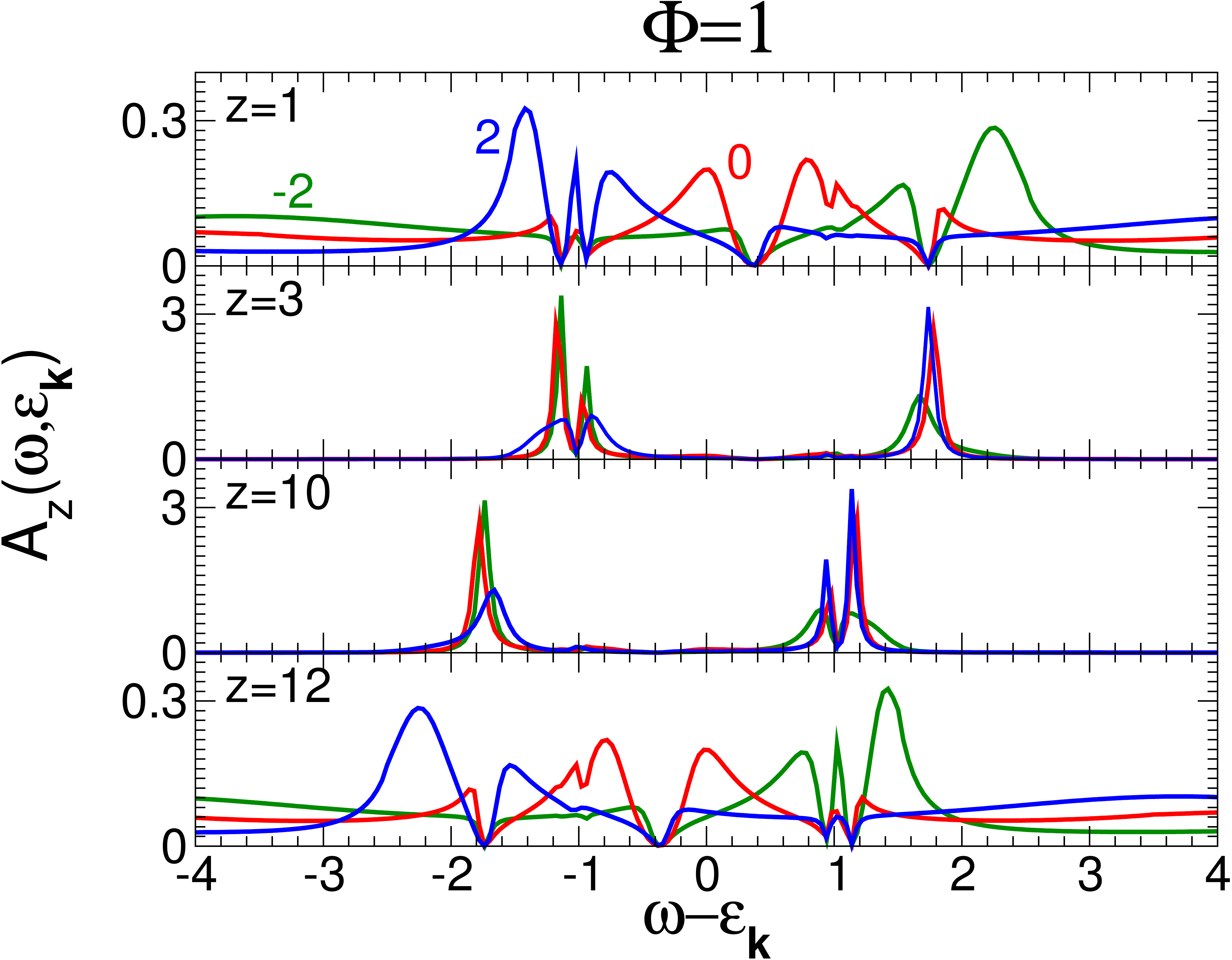}
\end{minipage}}
\hspace{0.025\textwidth}
\subfigure[]{
\label{Phi2.5}
\begin{minipage}[b]{0.45\textwidth}
\centering \includegraphics[width=1\textwidth]{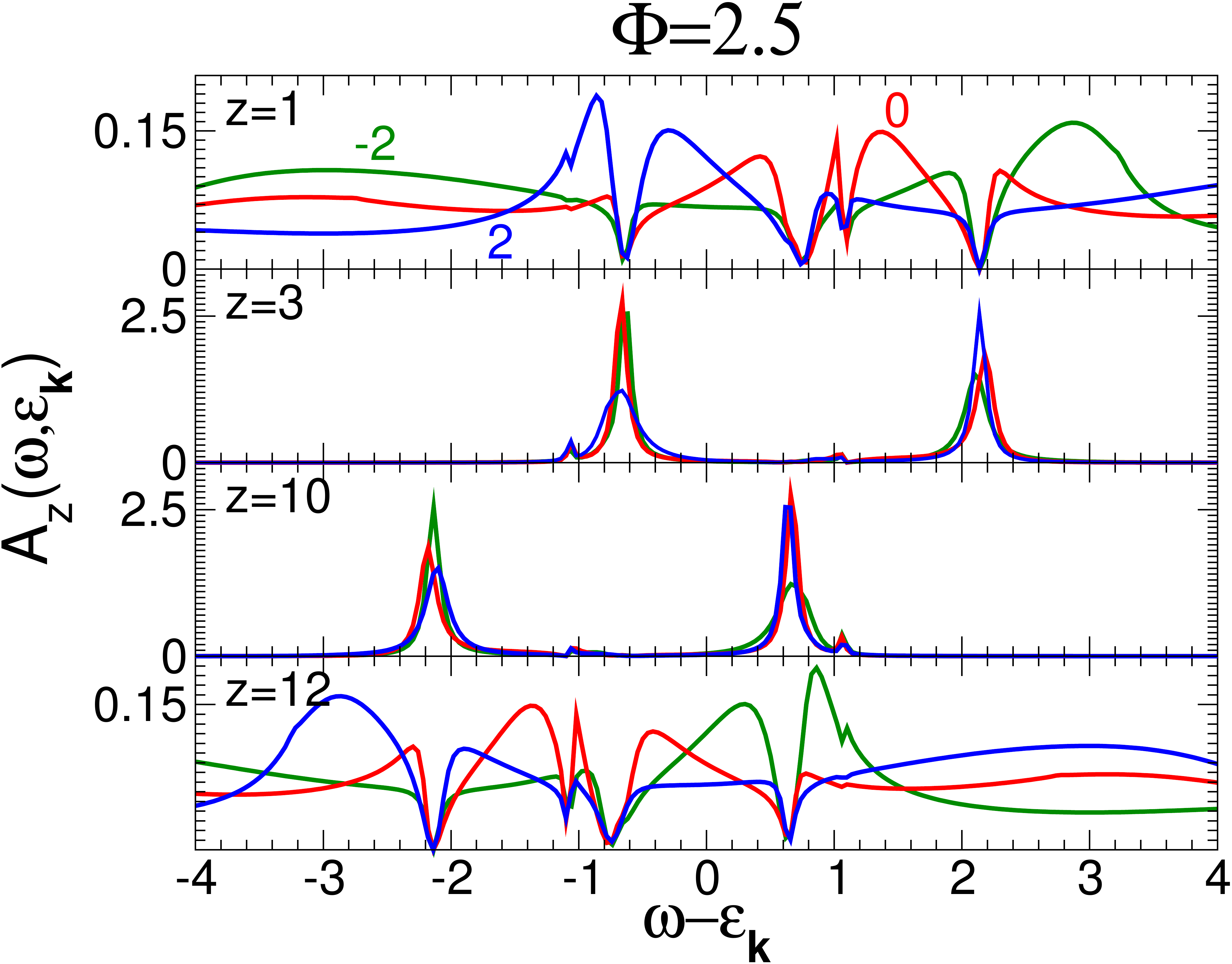}
\end{minipage}
}
\subfigure[]{
\label{Phi4}
\begin{minipage}[b]{0.45\textwidth}
\centering \includegraphics[width=1\textwidth]{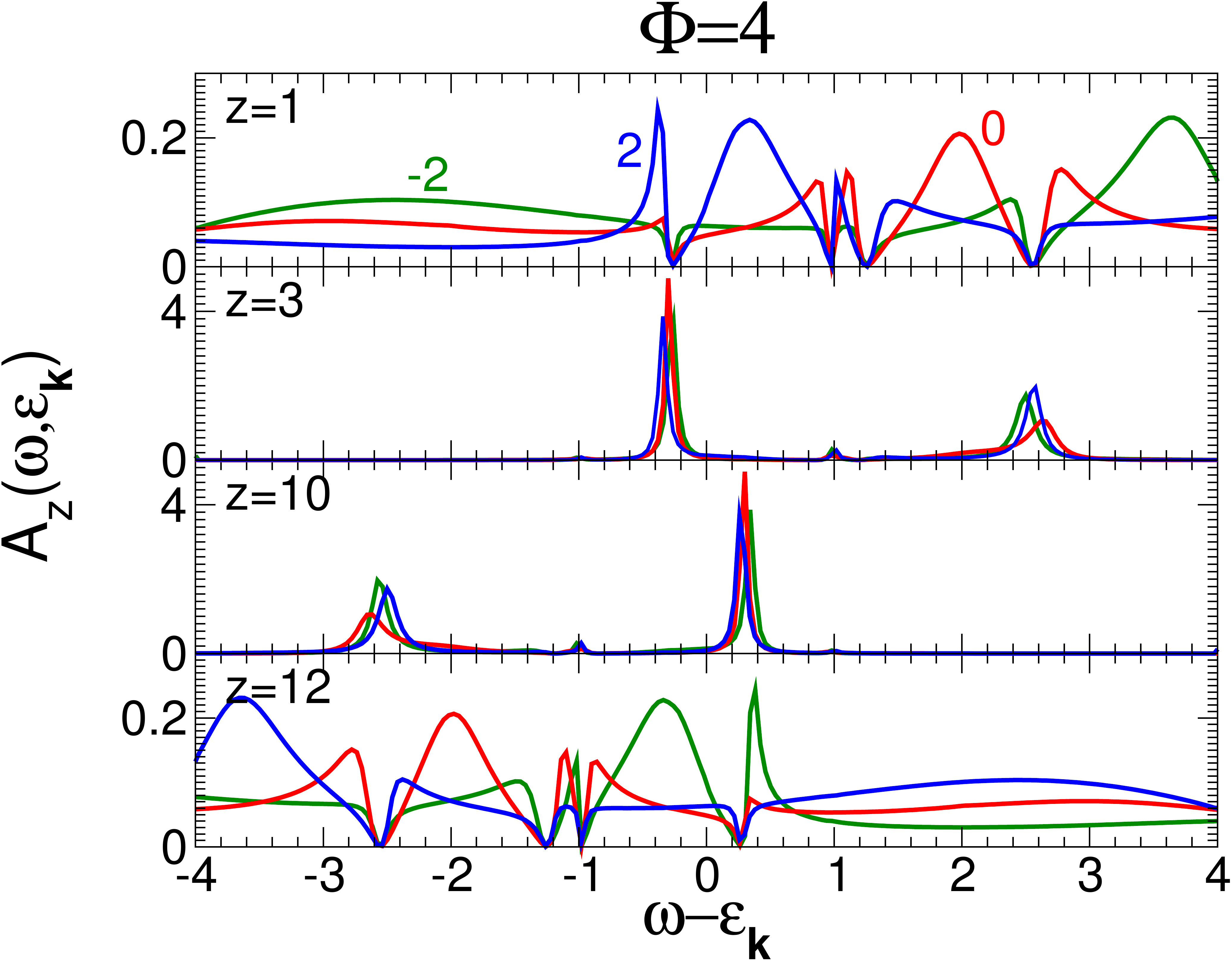}
\end{minipage}}
\hspace{0.025\textwidth}
\subfigure[]{
\label{Phi5.25}
\begin{minipage}[b]{0.45\textwidth}
\centering \includegraphics[width=1\textwidth]{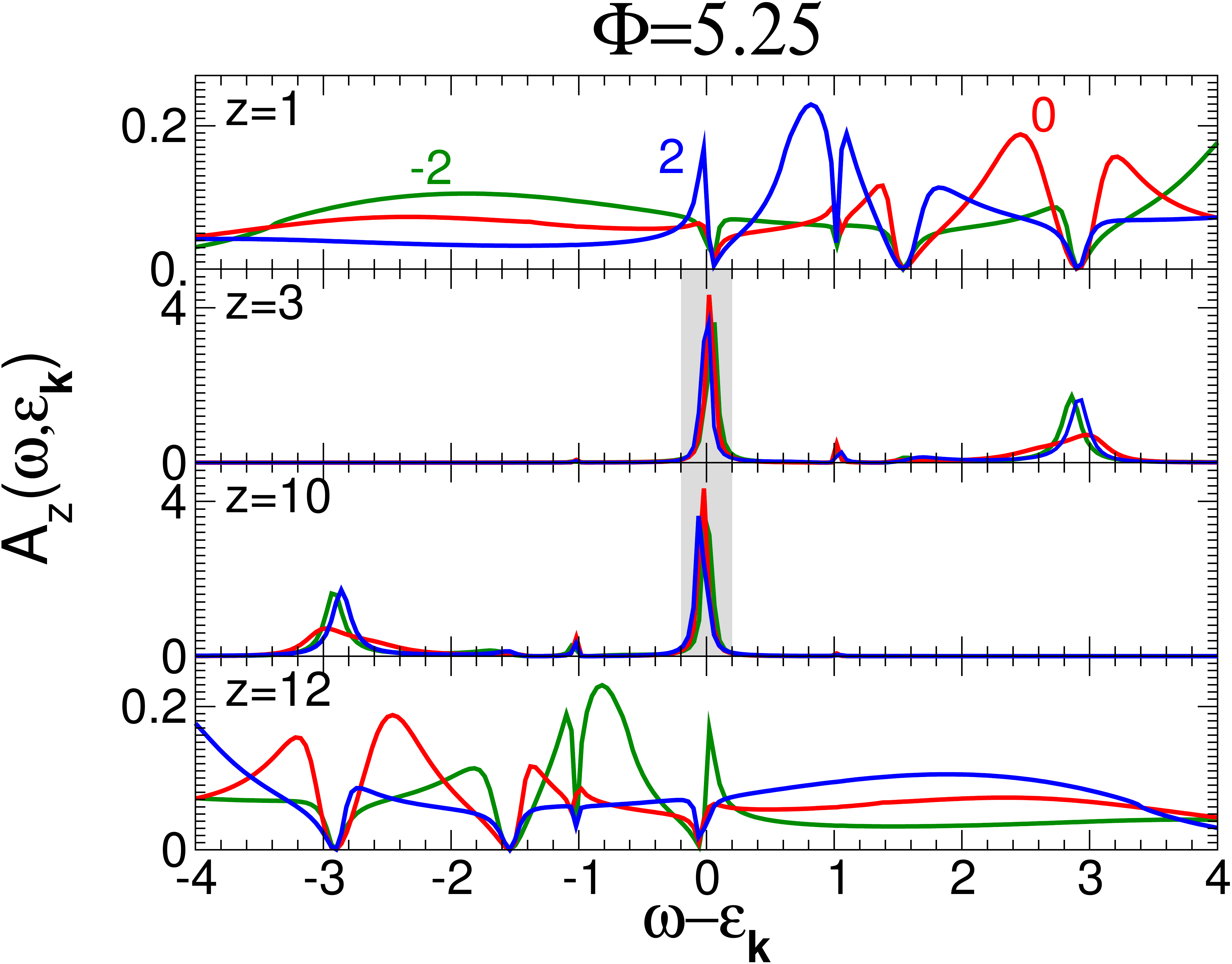}
\end{minipage}
}
\\
\caption{(Color online) Steady state spectral function $A_z(\omega, \varepsilon_{\kxy})$ for different values of bias voltage $\Phi$ and $\varepsilon(\kxy)$. In order to illustrate the resonance 
effect, we present results for bias voltages for which the current displays a maximum ($\Phi \simeq 2.5$ and $\Phi \simeq 5.25$),  a minimum ($\Phi \simeq 4$), and for a value in between ($\Phi=1$). The shaded 
area emphasizes the fact that for $\Phi \simeq 5.25$ the peak maxima of layers $z=3$ and $z=10$ overlap. 
Results are obtained with $N_b=6$. Here, $\varepsilon_{\kxy}=-2$ (Green), $\varepsilon_{\kxy}=0$ (red) and $\varepsilon_{\kxy}=2$ (blue). Other parameters are the same as in Fig.~\ref{schematicp} .}
\label{G_omega_epsilon}
\end{figure*}
\end{center}

\begin{figure}[t]
\includegraphics[width=0.75\columnwidth]{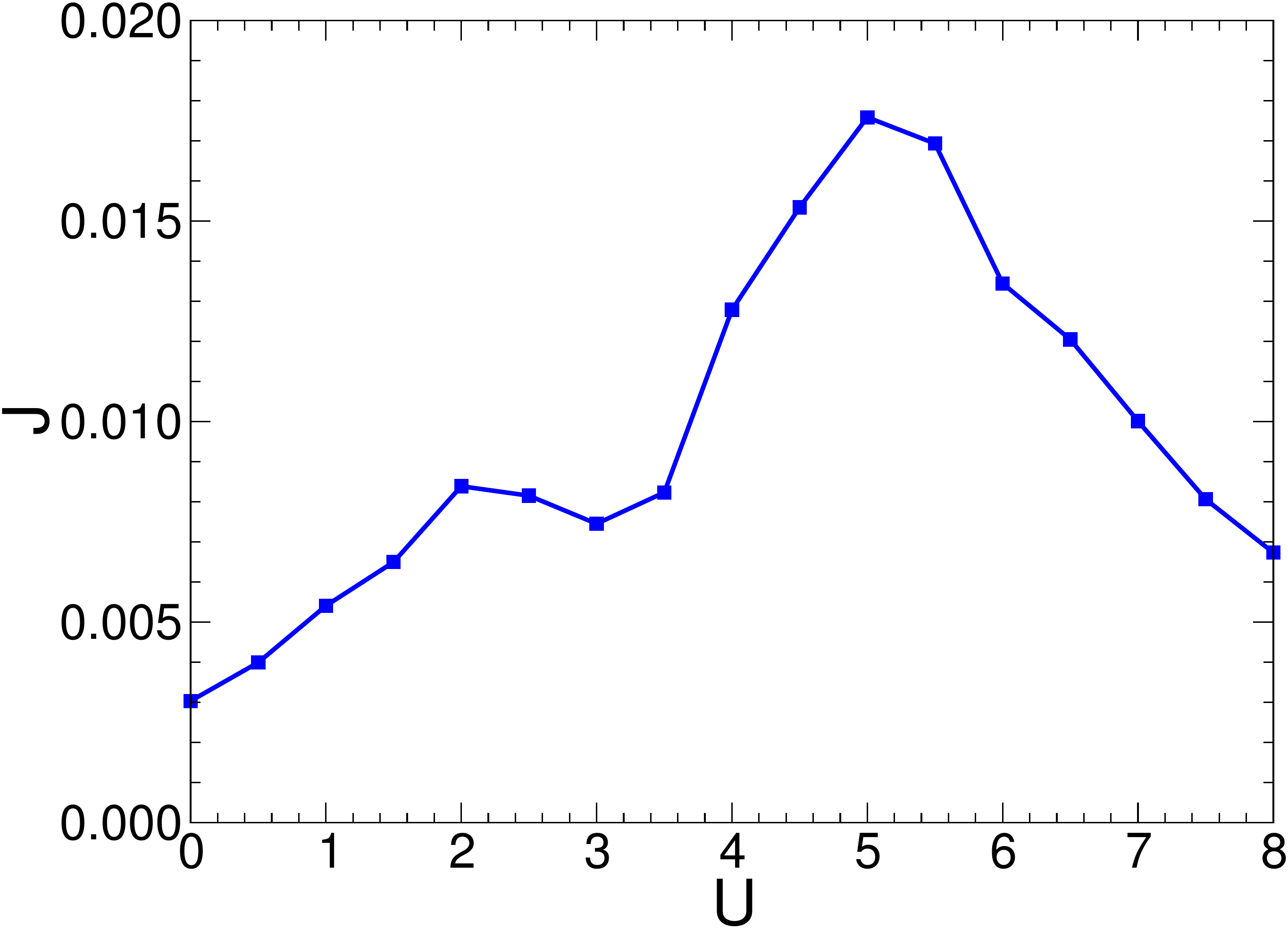}
\caption{Current $J$ as a function of the Hubbard interaction $U$ at the resonance. On-site energies in the first and the last layers are fixed to $\varepsilon_1^{(0)}=\varepsilon_{12}^{(0)}=-4$. 
Results are obtained with $N_b=4$. Other parameters are the same as in Fig.~\ref{schematicp} .}
\label{fig:J_vs_U}
\end{figure}

\section{Results}\label{Results}

Here, we presents results for the steady state properties of the system, displayed in Fig.~\ref{schematicp}, consisting of twelve layers (central region) sandwiched between two semi-infinite metallic leads.
Among these twelve  central region layers only the first and the last layers are correlated, with Hubbard interactions $U_1=U_{12}=U=8$ and onsite energies $\varepsilon_1^{(0)}=\varepsilon_{12}^{(0)}=-U/2=-4$. 
The onsite energies of the fifth and the eight layers are $\varepsilon_8^{(0)}=-\varepsilon_5^{(0)}=4$ and $\varepsilon_z^{(0)}=0$ for all $z \neq 1, 5, 8, 12$. The hopping between nearest-neighbor  central 
region sites $t_c=1$ is taken as unit of energy,\cite{initial_Sigma} while hopping between nearest-neighbor sites of the leads are $t_l=t_r=2$. Finally, the  
hybridizations between leads and  central region are $v_l=v_r=1$. All calculations are performed for zero temperature in the leads ($T_l=T_r=0$).  

The system is particle-hole symmetric. More specifically it is invariant under a simultaneous particle-hole transformation, a change of sign in the phase of one sublattice (as in the Hubbard model) together with 
a reflection of the $z$ axis. Therefore, properties of $z$-th and ${(L+1-z)}$-th layers are connected by  particle-hole transformation. Consequently, we need to evaluate the self-energy  for the $z=1$ layer only, 
and determine its value for $z=L$ layer based on the symmetry ($\Sigma_{12}^R(\omega)=-[\Sigma_{1}^R(-\omega)]^*+U$ and $\Sigma_{12}^K(\omega)=[\Sigma_{1}^K(-\omega)]^*$). All other layers are non interacting.

In Fig.~\ref{Current_vs_Phi} we plot the current-voltage characteristics of the system. Results are obtained with $N_b=2,4,6$ bath sites of the DMFT auxiliary impurity problem. We find that the difference between results 
obtained with $N_b=4$ and $N_b=6$ is small for all bias voltages. It indicates fast convergence of the current with respect to the bath sites $N_b$.

The current increases with increasing  bias voltage and reaches a first maximum at $\Phi \simeq 2.5$. Further increasing $\Phi$ reduces the current  until a minimum at $\Phi \simeq 4$ is reached.
A second maximum occurs at  $\Phi \simeq 5.25$. For larger bias voltages, the current again decreases due to the decreased  overlap of the density of states.

For low bias, where there is a large overlap of the density of states
of the left and the right leads, the conductivity is large and the
system is  in a high-conductivity regime.  That is why results in this
region are 
similar to the one of a single layer (see e.g. Refs. \onlinecite{ar.kn.13, ti.do.15}). In contrast, for larger bias $\Phi \gtrsim 3$  we are in the tunneling regime and the behavior of the current-voltage 
characteristics is significantly different.  As we discuss below, the
results we are showing are due to the occurrence of resonant tunneling.  To clarify this effect, we investigate the 
non-equilibrium spectral functions, which can be calculated from the corresponding Green's functions via $A_z(\omega, \varepsilon_{\kxy})=-\frac{1}{\pi}{\rm Im}G_z(\omega, \varepsilon_{\kxy})$.  
Due to the  geometry of the system (see fig.~\ref{schematicp}) three wells are formed in the intervals $2\le  z \le 4$, $6\le  z \le 7$, and $9\le  z \le 11$, to which electrons are partially confined and 
form quasi-bound levels. This can be seen by the fact that all spectral functions $A_z(\omega, \varepsilon_{\kxy})$ within a given well display peaks  for the same $(\omega, \varepsilon_{\kxy})$, 
corresponding to quantized quasi-stationary levels in this well. Electrons can leak from the one to the next well only by quantum tunneling.

In Fig.~\ref{G_omega_epsilon} we plot the steady state spectral functions $A_z(\omega, \varepsilon_{\kxy})$ as a function of $\omega-\varepsilon_{\kxy}$ for different $\varepsilon_{\kxy}$ and bias voltages $\Phi$. 
In particular, we show results for bias voltages that correspond  to maxima ($\Phi \simeq 2.5$ and $\Phi \simeq 5.25$), to a minimum and for a value ($\Phi=1$) at half maximum of the first peak in 
Fig.~\ref{G_omega_epsilon}.

The results have the correct property $A_{L+1-z}(\omega,\varepsilon_{\kxy})=A_{z}(-\omega,-\varepsilon_{\kxy})$, which is a consequence of the particle-hole symmetry of the Hamiltonian.
Our calculations show that for each non-correlated layer ($1<z<12$) the position of the peaks of the spectral function  
$A_z(\omega, \varepsilon_{\kxy})$ depends only on the value of $\omega-\varepsilon_{\kxy}$ and not on $\omega$ and $\varepsilon_{\kxy}$ separately. This indicates that for the 
non-correlated layers one-dimensional physics dominates and $\varepsilon_{\kxy}$ only shifts the energy levels.
Furthermore, peaks of the spectral functions $A_z(\omega, \varepsilon_{\kxy})$ for the non-correlated layers in the first ($z=2,3,4$) and the last ($z=9,10,11$) wells generate dips in the spectral 
functions $A_z(\omega, \varepsilon_{\kxy})$ of the first ($z=1$) and the last ($z=12$) correlated layers correspondingly. This can be qualitatively understood from  Eq.~\eqref{GRzz}, if one assumes that 
$[{\bf G^{-1}}]_{zz}^R$ is a smooth, function, while $-L_{z-1}^R$ or $-R_{z+1}^R$ (neighboring layer Green's functions) have narrow peaks.

As central region (layers $1<z<12$) are non-interacting
resonant tunneling occurs when quasistationary states, i.e. the peaks in the spectral function, of the first and the last well coincide for any $\varepsilon_{\kxy}$.\cite{er.po.11, gu.to.16} This is the case for 
$\Phi \simeq 5.25$, as can be seen by the gray regions in Fig.~\ref{Phi5.25}. If these peaks are within the energetic transport window the current gets enhanced at the corresponding bias voltage. For all other 
bias voltages shown (see Figs.~\ref{Phi1}-\ref{Phi4}), peaks of $A_z(\omega, \varepsilon_{\kxy})$ for  different wells do not coincide, so no resonant tunneling takes place. The second maximum in the 
current-voltage characteristics (see Fig.~\ref{Current_vs_Phi}) can, therefore, be understood in terms of such a resonant tunneling effect. On the other hand, the first maximum is due to the finite bandwidth of 
the leads, similar to the one for a single layer case (see e.g. Refs. \onlinecite{ar.kn.13, ti.do.15}). In contrast to the single layer case, in the current situation electrons tunnel through four layers 
($z=1,5,8,12$) and therefore the current drops faster after the maximum.

In order to address the effect of electron correlation on the resonance, we investigate the behavior of the resonance current $J$ as a  function of the interaction $U$.~\cite{barrier} In Fig.~\ref{fig:J_vs_U} 
we plot the current $J$ as a function of the interaction $U$ at the corresponding resonance bias voltage. The figure clearly shows that correlation effects substantially enhance the resonance effect. However 
the current maximum is obtained at not too large values of $U\sim 5$. This enhancement  behavior can be understood in terms of  two competing effects occurring as a function of $U$: since the  resonance takes 
place at relatively high bias, the {\it one-dimensional} density of states (DOS) of the two leads have a reduced overlap. This suppresses tunneling at small $U$ for which scattering (approximately) conserves 
the momentum parallel to the layers. Upon increasing $U$, scattering channels to different values of the in-plane  {\bf k}  open, so that the {\it three-dimensional} DOS is available for scattering, thus 
enhancing the current. On the other hand, by increasing $U$ also backscattering is increased, which, in turns suppresses the current.

\section{Conclusions}\label{Conclusions}

Using non-equilibrium DMFT calculations we investigate steady state properties of a multilayer heterostructure consisting of correlated and non-correlated layers. Due to the fact that the system is inhomogeneous, 
no matter how many impurity problems have to be solved, ``standard'' DMFT is not applicable and one has to use the real-space generalization of it. As an impurity solver we used the recently introduced auxiliary 
master equation approach, which addresses the impurity problem within an auxiliary system consisting of a correlated impurity, a small number of uncorrelated bath sites and two Markovian environments described 
by a generalized master equation\cite{ar.kn.13, do.nu.14, ti.do.15}. 

In particular, our main goal was to investigate resonance effects in this system. For this purpose we chose an arrangement of layers such that electrons were confined in three different wells and transport 
through the central region was only possible by quantum tunneling. For a particular bias voltage ($\Phi \simeq 5.25$) we observed that quasi-stationary energy levels  in the first and the last wells 
coincided and resonance tunneling between them takes place. At that bias voltage the current displays a maximum. According to our calculations the current has  another maximum at $\Phi \simeq 2.5$. The latter 
is due to the finite bandwidth of the leads.  We checked that these qualitative findings are robust up to some extent as a function of the model parameters.

Furthermore, we also investigate effect of the interaction strength on the current at the resonance.  We obtain that correlation effects for weak up to strong interaction substantially enhance the resonance
current.

\begin{acknowledgments}
We thank Andreas Weichselbaum, Jim Freericks, Elias Assmann, Max Sorantin for valuable discussions. 
This work was supported by the Austrian Science Fund (FWF):  P26508, as well as SfB-ViCoM project F04103, and NaWi Graz. The calculations were partly performed on the D-Cluster Graz 
and on the VSC-3 cluster Vienna. 
\end{acknowledgments}

\bibliographystyle{aipnum4-1}

\begin{thebibliography}{97}
\bibitem [{oka()}]{okam}%
  \BibitemOpen
  \href@noop {} {}\bibinfo {note} {For a treatment of a similar nonequilibrium
  system by R-DMFT see also ~\cite{okam.08}. There a NCA impurity solver was
  used.}\BibitemShut {Stop}%
\bibitem [{con()}]{convergance}%
  \BibitemOpen
  \href@noop {} {}\bibinfo {note} {We compare two consecutive hybridization
  functions in the DMFT iteration and use as convergence criterion that the
  root-mean-square deviation \hbox{$\sqrt{\int d\omega\left(|\Delta_{\rm
  new}^R(\omega)-\Delta_{\rm old}^R(\omega)|^2+ |\Delta_{\rm
  new}^K(\omega)-\Delta_{\rm old}^K(\omega)|^2\right)}<10^{-6}$}. Here we also
  note that $\int d\omega|\Delta^\gamma(\omega)|>1$.}\BibitemShut {Stop}%
\bibitem [{ini()}]{initial_Sigma}%
  \BibitemOpen
  \href@noop {} {}\bibinfo {note} {Usually, hopping $t_c$ is of the order of
  1eV.}\BibitemShut {Stop}%
\bibitem [{bar()}]{barrier}%
  \BibitemOpen
  \href@noop {} {}\bibinfo {note} {These results are obtained by
  fixing a potential barrier $\varepsilon_1^{(0)}=\varepsilon_{12}^{(0)}=-4$
  for all $U$ values, otherwise there would be no barrier at
  $U=0$.}\BibitemShut {Stop}%
\end{thebibliography}

%

\end{document}